\documentclass[unnumsec,webpdf,modern,large]{mam-authoring-template}%

\usepackage[version=4]{mhchem}

\graphicspath{{figures/}}

\theoremstyle{thmstyleone}%
\theoremstyle{thmstyletwo}%
\theoremstyle{thmstylethree}%

\begin{document}

\journaltitle{Microscopy and Microanalysis}
\DOI{DOI HERE}
\copyrightyear{2025}
\pubyear{2025}
\access{Advance Access Publication Date: Day Month Year}
\appnotes{Original Article}

\firstpage{1}


\title[Fast phase mapping in 4D-STEM]{Fast 4D-STEM-based phase mapping for amorphous and mixed materials}

\author[1,]{Andreas Werbrouck\ORCID{0000-0003-3796-0024}}
\author[2]{Nikhila C. Paranamana\ORCID{0000-0001-9466-5780}}
\author[3]{Xiaoqing He\ORCID{0000-0001-6058-9732}}
\author[1,2,4]{Matthias J. Young$\ast$\ORCID{0000-0001-7384-4333}}

\authormark{Werbrouck et al.}

\address[1]{\orgname{University of Missouri Materials Science and Engineering Institute}, \orgaddress{\postcode{65201}, \state{Columbia, MO}, \country{USA}}}
\address[2]{\orgdiv{Department of Chemistry}, \orgname{University of Missouri}, \orgaddress{\postcode{65201}, \state{Columbia, MO}, \country{USA}}}
\address[3]{\orgname{University of Missouri Electron Microscopy Core}, \orgaddress{\postcode{65201}, \state{Columbia, MO}, \country{USA}}. Currently at \orgname{Institute of Materials Science and Engineering, Washington University in Saint Louis}, \orgaddress{\state{Saint Louis, MO}, \country{USA}}}
\address[4]{\orgdiv{Department of Chemical and Biomedical Engineering}, \orgname{University of Missouri}, \orgaddress{\postcode{65201}, \state{Columbia, MO}, \country{USA}}}

\corresp[$\ast$]{Corresponding author. \href{email:matthias.young@missouri.edu}{matthias.young@missouri.edu}}

\received{Date}{0}{Year}
\revised{Date}{0}{Year}
\accepted{Date}{0}{Year}

\abstract{Interpretation and mapping strategies for 4D-scanning transmission electron microscopy (4D-STEM) are well-developed for crystalline materials, yet in the case of amorphous and mixed materials it is significantly more challenging to separate different phases. Non-negative matrix factorization (NMF) in principle would allow separation of 4D-STEM data into components with interpretable diffraction signatures and intensity maps, independent of the crystalline, amorphous or mixed nature of the material.
However, adoption of NMF in this field is hampered by large datasets and conceptual hurdles: NMF tackles a non-convex optimization problem, requiring iterative algorithms. Additionally, the stopping condition has to be chosen carefully. In this work, we show that the factorization of large 4D-STEM datasets can be drastically accelerated using a QB decomposition (i.e. randomized NMF or RNMF), leading to much shorter time per iteration. This allows structure-independent phase mapping on very large 4D-STEM datasets. 
We validate this approach on a synthetic literature dataset (mixed \ce{ZrCuAl}), before mapping a thin \ce{TiO2} layer on top of \ce{SiO2}, and an interface between a lithium-ion cathode and solid state electrolyte. We also demonstrate that, before using NMF to transform the data on an interpretable, nonnegative basis, principal component analysis (PCA) can be used for fast exploratory analysis to assess dataset dimensionality and linearity. 
}




\maketitle

\section{Introduction}

\begin{figure}[!t]
\centering
\includegraphics[width=\columnwidth]{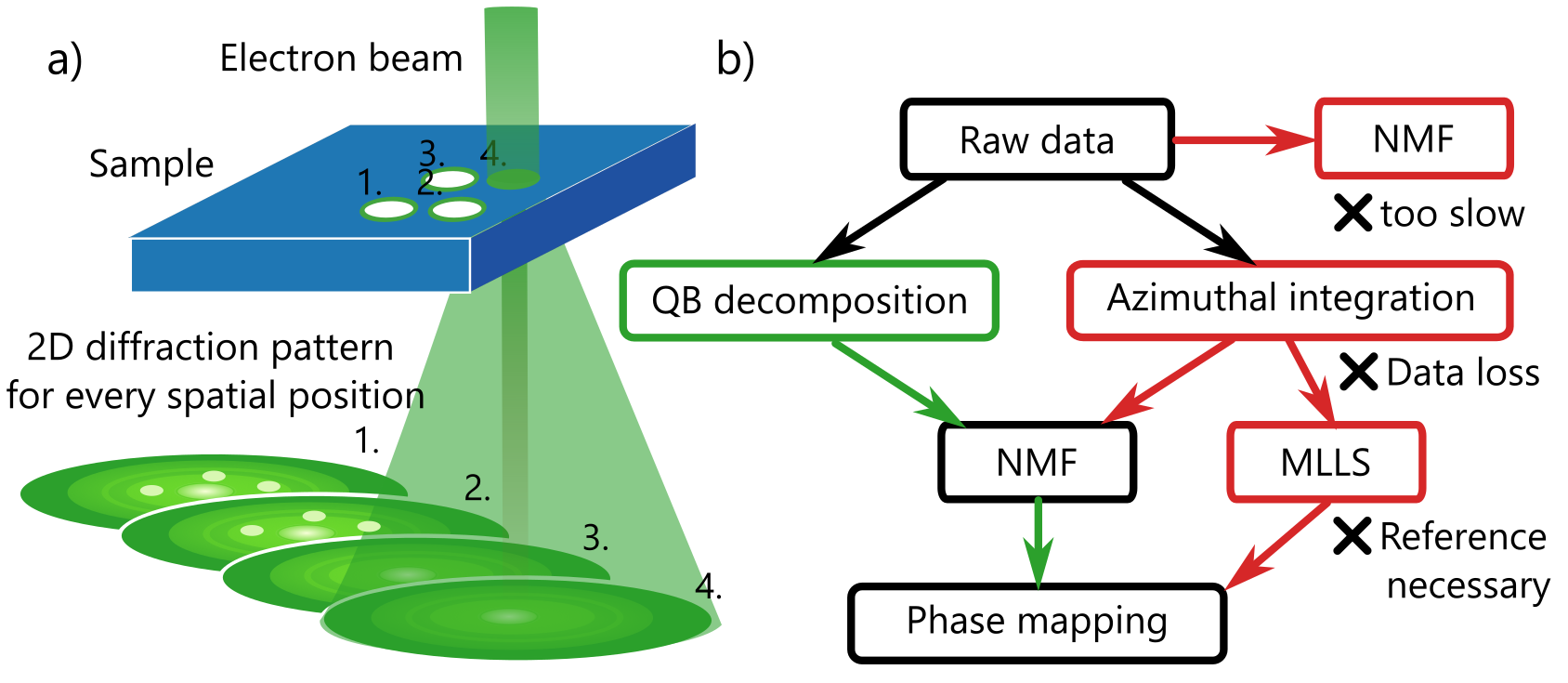}
\caption{a) 4D-STEM working principle. b) Dataflow diagram of our approach (green), in contrast with other methods (red).}
\label{fig_0}
\end{figure}

The ability to engineer materials and interfaces on an atomic scale has been pivotal in technological advances for semiconductors \citep{plummer2023integrated}, catalysis \citep{polshettiwar2010green}, nanomedicine \citep{riehemann2009nanomedicine}, and energy storage \citep{pomerantseva2019energy}. Transmission electron microscopy (TEM) provides the ability to directly image these structures, giving tremendous insight into what is happening at the nanoscale \citep{Williams2009Transmission}.  When a TEM instrument is used in scanning transmission electron microscopy (STEM) mode, the electron beam is focused onto a small area and scanned over the sample to collect an image. In recent years, great strides have been made in the development of fast large area detectors \citep{Booth2012K2, Tate2016High}, further expanding the boundaries of what is experimentally possible. One of these novel techniques is 4D-STEM, in which the scanning beam is used to collect a 2D diffraction micrograph for every scanned spot \citep{Ophus2019Four} (Figure \ref{fig_0}a). 

This effectively creates a four-dimensional dataset: two dimensions for the real space (positions on the sample), while the two remaining dimensions contain information about the reciprocal space (distances between atoms and atomic planes) for every position. The enormous flexibility to collect data in real and reciprocal space at high acquisition rates makes 4D-STEM a versatile and exciting tool for materials research.

Each 2D diffractogram within the 4D-STEM data contains structural information about the material in the beam at that position. Crystalline materials produce sharp Bragg peaks from elastic scattering of atomic planes. These can be indexed to determine the crystal structure. Mapping crystalline structures over a large amount of measurements is relatively straightforward by defining a `virtual detector', plotting the variation in intensity of the signal of a specific spot on the detector where one expects a Bragg peak. Another approach for mapping orientation and/or strain in crystalline structures is Automated Crystal Orientation Mapping (ACOM) \citep{ophus2022automated}. 

In contrast to crystalline materials, elastic scattering of an electron beam off atoms in amorphous phases leads to diffuse ring-like 2D patterns, which, after azimuthal integration, can be transformed into a reduced pair distribution function (PDF) G(r) \citep{Egami2012Underneath} that provides real-space atomic structure information about the amorphous material. Due to the continuous nature of this signal, it is difficult to map different phases using the same virtual detector approach.

Previous work has presented strategies to automate the analysis of large 4D-STEM datasets (red portions of Figure \ref{fig_0}b): Mu \textit{et al.} performed a hyperspectral analysis on a 4D-STEM dataset of amorphous and semicrystalline organic thin films \citep{Mu2019Mapping}. The reciprocal space data were azimuthally integrated, and a combination of reference radial distribution functions (functionally equivalent to PDFs) was fitted to each pixel with multiple linear least squares (MLLS). The same method was used by Donohue \textit{et al.} \citep{Donohue2022Cryogenic} to map amorphous-crystalline blends. Such a reference RDF can be seen as an effective `virtual detector' for amorphous materials. However, finding appropriate reference RDFs is an additional hurdle to performing such analysis and is limited to one-dimensional PDF analysis of amorphous or polycrystalline phases. 

One of the implicit assumptions of employing reference RDFs to fit PDF data is that under the kinematic (single-scattering) approximation, diffraction intensities from a mixed-phase material superpose linearly: if a beam position samples a mixture of two phases, the resulting diffraction pattern is assumed to be a weighted sum of the individual phase patterns. 

Finding and describing these \textit{linear} subspaces is the goal of blind source separation in hyperspectral unmixing \citep{gillis2020nonnegative}. In this domain, work has been done with matrix decomposition techniques such as principal component analysis (PCA) and non-negative matrix factorization (NMF).  

PCA and NMF have been used in the past for 4D-STEM mapping \citep{Allen2021Fast}\citep{Uesugi2021Non}\citep{Donohue2022Cryogenic}\citep{Kang2024Mapping}\citep{Kimoto2024Unsupervised}\citep{kimoto2025nonnegative}. In this work, we recognize that NMF offers a meaningful solution to the problem of blind hyperspectral unmixing in 4D-STEM. It does not need reference spectra and, in contrast to PCA, yields interpretable components. NMF relies on iterative algorithms, which has led to two issues hampering its wider adoption: 4D-STEM datasets are too large and convergence is poorly defined.

\textbf{Very large datasets.} ideally one wants to use 2D diffraction spectra as opposed to compressed (e.g. azimuthally integrated) spectra to retain as much interpretability as possible. Execution time per iteration scales with the number of positions probed, the number of pixels in the diffraction detector and the number of components. As such, large datasets lead to \textit{slower iterations}. 

\textbf{Initialization and convergence criteria.} Different initialization schemes for NMF have been developed over the years \citep{Boutsidis2008SVD, gillis2013fast}. Random initialization will converge to \textit{a} local minimum. But convergence can be reached faster starting from an approximate factorization, such as non-negative double singular value decomposition (NNDSVD). 
It is computationally rather expensive to calculate the difference between data and model every iteration. Hence, projected gradients are used instead. These gradients are normalized with respect to the initial gradient, which in turn depends on the initialization scheme. 
This leads to a situation in which \textit{too many expensive iterations} are performed, with the scaling of the convergence criterion depending on the initialization. However, if the number of iterations is fixed \textit{a priori} or dictated by computational constraints, the solution may not have converged when iterations are stopped. 

In this work, we propose to use QB decomposition, a low-rank matrix approximation technique, as a preprocessing step for NMF to perform \textit{Randomized} NMF (RNMF) \citep{Erichson2018Randomized}. QB decomposition samples the linear subspace efficiently, effectively compressing measurement (or feature) space. In this way, the processing bottleneck is drastically reduced. This reduction in size of the dataset enables running many more iterations in a shorter time, making sure convergence is reached.

\section{Methods}
\subsection{Data representation}
Consider a diffraction dataset stored in a 4D matrix $X_{m_xm_yn_xn_y}$ consisting of $m = m_x \times m_y$ spatial pixels, where for every spatial pixel there is a 2D diffraction measurement with $n_x$ horizontal pixels and $n_y$ vertical pixels. This 4D matrix can be reshaped into a 2D  matrix $X_{mn}$ for analysis: the 2D spatial pixels are stacked below each other in $m = m_x \times m_y$ rows. Each diffraction pattern is unrolled into a row of length $n = n_x \times n_y$. As such, we have reimagined the 4D dataset $X_{m_xm_yn_xn_y}$ as a 2D dataset $X_{mn}$ with $m$ samples and $n$ features. This approach is identical to that described in \citep{Kimoto2024Unsupervised} and other work.

If every measured count originates from a physical, linear process, and there are k such processes contributing to the measurement, the n data points in this dataset lie approximately in a k-dimensional subspace in m-dimensional space (a convex cone or, more strictly, a simplicial cone). In other words, measurement noise nothwithstanding, the matrix $X_{mn}$ has rank $k$. 

While there are theoretically infinitely many possible $k$-dimensional bases into which this matrix can be projected, the \textit{number} of unique components necessary to accurately reconstruct the matrix should be independent of the basis selected, and should correspond to the number of superposed physical, linear processes occurring in the measured data. Principal component analysis and nonnegative matrix factorization are two methods for projecting a matrix into a $k$-dimensional subspace. While the base components derived from these methods will be distinct, the dimensionality (value of $k$) is expected to be the same for both methods. 

\subsection{Principal component analysis}
Principal component analysis (PCA) is a linear dimensionality reduction technique that decomposes a data matrix into a set of orthogonal components, describing the directions of greatest variance. Given $X_{mn}$, PCA decomposes the data as

\begin{equation}
    X_{mn} = \mu_n + \sum_{i=1}^{k}{W_{mi}H_{in}}
\end{equation}

where $\mu_n$ is the mean of all $m$ rows to center the decomposition on the mean. Due to this centering on the mean, $W_\text{pca}$ and $H_\text{pca}$ necessarily contain negative elements, reducing interpretability. However, PCA offers a computationally efficient way to describe the linear subspace in which the data are located. 

\subsection{Nonnegative matrix factorization}
NMF \citep{Lee1999Learning} separates $X_{mn}$ into $k$ non-negative vectors $\mathbf{W}$ ($m \times k$ matrix) and $k$ non-negative vectors $\mathbf{H}$ ($k \times n$ matrix):

\begin{equation}
    X_{mn}=\sum_{i=1}^{k}{W_{mi}H_{in}}
\end{equation}

Here, the $k$ rows of $H$ can be interpreted as fundamentally distilled `fingerprint diffraction spectra' of the present species with a high signal-to-noise ratio, but without any spatial dependency. Consequently, the $k$ columns of $W$ can be interpreted as the spatial distribution of these species, mapping the different components in the measured region. 

Hence, NMF differs from PCA in two ways: the subspace is anchored at the origin (no mean term $\mu_m$), and the basis vectors $W_i$ are constrained to be non-negative (rather than constructed along the axes of greatest variation). If the data are indeed generated by $k$ linear, non-negative physical processes, the corresponding mixtures lie in a $k$-dimensional (simplicial) cone. Because the dimensionality (value of $k$) is expected to be independent of the basis selected, one can employ PCA explained variance as an efficient means to determine the dimensionality of the non-negative physical processes, and subsequently perform NMF with that number of components to obtain a physically interpretable non-negative factorization.

\begin{figure}[!t]
\centering
\includegraphics[width=\columnwidth]{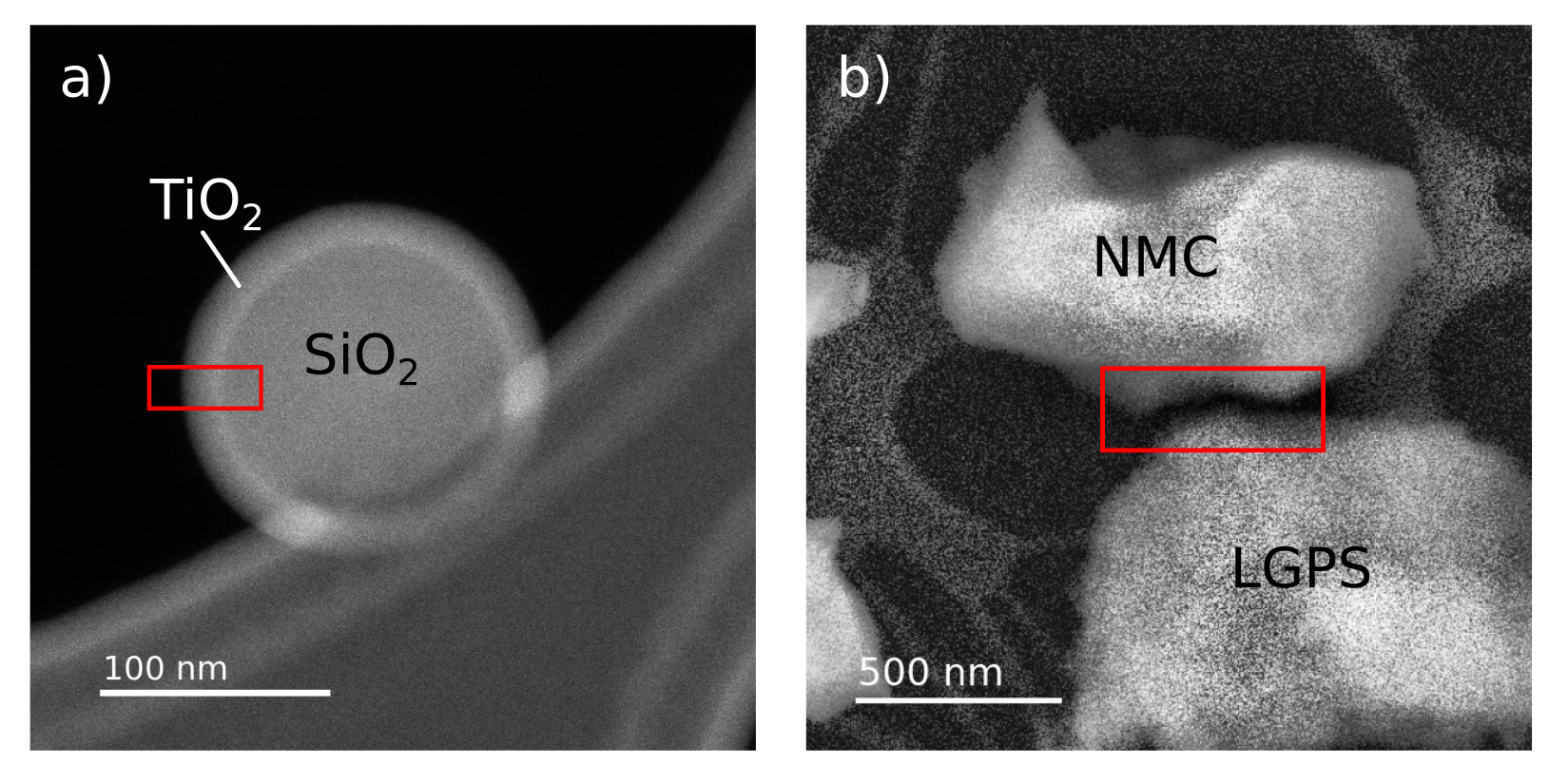}
\caption{High-Angle Annular Dark-Field (HAADF)-STEM images of the example material systems used to benchmark RNMF of 4D-STEM data. a) \ce{TiO2} coating on a \ce{SiO2} sphere b) NMC-LGPS interface. 4D-STEM areas are indicated with red rectangles.}
\label{fig_1}
\end{figure}

Factorization into non-negative components is a non-convex problem, without a guaranteed global optimum. As a result, iterative algorithms have to be used. For this, the user makes choices regarding initialization, loss function, number of components, iterative algorithm, and stopping criterion. Below, we add a non-standard data preprocessing step to reduce the time per iteration and we discuss our post-processing and visualization approach. Extensive literature is available on the topic of NMF \citep{gillis2020nonnegative}, blind source separation \citep{choi2005blind} and hyperspectral unmixing \citep{feng2022hyperspectral}, \citep{heylen2014review}. In the following, we summarize some key points relevant for the context of 4D-STEM. 

\textbf{Initialization}. The most general and robust initialization is to fill W and H with scaled random numbers. Several other initialization schemes have been developed, such as the Successive Projections Algorithm (SPA) \citep{gillis2013fast}. The initialization we will use is the Non-negative singular value decomposition (NNDSVD) as developed by \cite{Boutsidis2008SVD}. In contrast with random initialization, NNDSVD provides a \textit{very good} first guess for W and H. Successive iterations then refine these initial estimates, but the heavy lifting is done by the initialization. In this work, we use NNDSVDar, a small modification of NNDSVD in which zero components are filled with small random variables. 

\textbf{Stopping criterion}. Because of the additional non-negativity constraint, it is generally not possible for NMF to reach the global minimum in the error landscape as defined by PCA. Hence, convergence to a local minimum can be determined through iterative improvement of the output of the loss function. When the improvement decreases below a certain threshold, the decomposition is considered good enough. Since the structure of the problem is unknown, the initial error is often used as a scaling factor. Although this is not wrong, it is harder to improve upon a very good initialization such as NNDSVD compared to improving a random initialization. Here, we suppose that 1000 iterations over our initial W and H determined with NNDSVDar are sufficient for convergence.  

\textbf{Data preprocessing}. Randomized non-negative matrix factorization (RNMF) was proposed by \citep{Erichson2018Randomized}, using concepts introduced by \citep{Halko2011Finding}. It was shown that $X_{mn}$ can be replaced with a smaller, $m \times (k + q)$ matrix $B$ with $k + q << n$, where $q$ is an oversampling parameter and $n$ is the largest dimension of the matrix (due to symmetry the matrix $X$ can be trivially transposed without consequence except swapping the meaning of W and H). The matrix $B$ can be constructed from X such that $X = QB$ with Q an orthogonal, `random' gaussian mixing matrix. This process is also known as QB decomposition \citep{Halko2011Finding}. B is then factorized into components $\widetilde{W}$ and $H$. The matrix $\widetilde{W}$ can be scaled back to $W \ =\ Q\widetilde{W}$, so we can factorize $X \ =\ QB \ =\ Q\widetilde{W} H \ =\ WH$. 


This means that the sample space remains the same, but the feature space is reduced massively (or vice versa), leading to scaling with $\sim$ O(mk) instead of $\sim$ O(nmk). 

Furthermore, the error introduced by this approximation can be quantified and is usually very small. Specifically for 4D-STEM datasets, either the number of pixels in the detector or the number of spatial positions can be compressed by the QB decomposition. The larger matrix dimension often be the reciprocal space dimension, but in the case of a pixelated detector, the larger dimension may also be the real space dimension. It is worth noting that the term \textit{randomized} non-negative matrix factorization concerns the random gaussian mixing matrix Q, and not the NMF initialization. When the term RNMF is used, we refer to NMF with QB decomposition preprocessing, in line with \citep{Erichson2018Randomized}.

\textbf{Post-processing}. Although, as mentioned before, W and H may not be unique, any given W and H are trivially degenerate up to a factor and can be scaled to yield the same results. Hence, normalizing each component H and scaling the corresponding map W accordingly allows one to spatially resolve and compare relative intensities of these components. We adopt this approach, as employed in previous work \citep{gillis2013fast} in the rest of the manuscript. 

\textbf{Visualization} The rows (and columns) of W (and H) are reorganized to their original 2D shapes. For each analysis, the absolute residual $r_{mn}=|X_{mn}-\Sigma_iW_{mi}H_{in}|$ is calculated. $r$ is also four-dimensional and has to be compressed to show average residuals in real space and reciprocal space, respectively, through $r_{real}=\max_{n\ } \left|r_{mn}\right|$ and $r_{reciprocal}=\max_{m\ }\left|r_{mn}\right|$. These compressed 1D vectors can then be reshaped into 2D maps to be interpreted. 
We are aware that the expression of $r$ carries some assumption about the data and the underlying noise distribution. As we show below, there is no one universally good solution for this, hence we believe the absolute value of the difference between model and data is a generally interpretable choice.

We found it useful to visualize the normalized $H$ matrices on a semi-logarithmic scale, in which the region 1-0.01 is shown logarithmically, and 0.01-0 on a linear subscale. Similarly, PCA components (in SI) are shown on a symmetric logarithmic scale with a linear section depending on the situation. All other colorscales are linear. We have strived to make these choices consistently and indicate them in a transparent way, with the main aim to show relevant features where possible. 


\subsection{Literature dataset}
The \ce{ZrCuAl} dataset \citep{kimoto2025nonnegative} was kindly provided by Dr. Koji Kimoto. It consisted of 32 $\times$ 32 (1024) positions in real space, and 128 $\times$ 128 (16384) pixels in reciprocal space. Each real space pixels received a dose of $10^4$ counts (with Poisson noise). 3 $\times$ 3 crystalline grains with the same orientation and various degrees of crystallization are embedded in an amorphous matrix, giving rise to 4 components (background and 3 crystal orientations). 

\subsection{Materials preparation}
\ce{SiO2} spheres (NanoXact, 120 nm diameter) were loaded onto a TEM grid and placed in the holder described in earlier work \citep{Jasim2021Cryo}. Atomic Layer Deposition (ALD) of \ce{TiO2} was conducted in a home-built tube reactor under 1 torr Ar, at 150°C. 400 cycles of \ce{TiCl4} (2s pulse, 30s purge) and \ce{H2O} (2s pulse, 30s purge) yielded a thickness of 23.1 nm (front) and 20.8 nm (end) on witness Si coupons. This thickness is within expectations for this chemistry \citep{Ritala1993Growth}.

An extensive description of the preparation of the NMC-LGPS dataset can be found in our earlier work \citep{Paranamana2025Understanding}.

\subsection{Transmission Electron Microscopy}
A Thermo Fisher aberration corrected Spectra 300 TEM with Ceta-M camera was used for both STEM diffraction and EDS measurements at 300 kV. STEM image and diffraction measurement were collected in the STEM microprobe mode using a spot size 6 with condenser lens apertures of C1, C2, and C3 being 2000 $\mu$m, 50 $\mu$m and 1000 $\mu$m respectively. The semi-convergence angle was set to 1.1 mrad. The nominal camera length for electron diffraction measurements was calibrated using a Si standard to be 37 mm. 

4D-STEM data for the first experimental dataset (\ce{TiO2} on \ce{SiO2} as depicted in Figure \ref{fig_1}a.) included 34969 features per diffraction pattern, and 2028 positions in real space. The second dataset described is the NMC-LGPS interface in Figure \ref{fig_1}b, which had 2268 positions in real space and 34969 features per diffraction pattern. 

\subsection{Computational approach}
The python implementation (scikit-learn) of NMF was used and adapted. RNMF uses the same implementation after QB decomposition, allowing for straightforward comparison. In all cases, non-negative double singular value decomposition with arithmetic rescaling (NNDSVDar) component initialization \citep{Boutsidis2008SVD} was followed by a coordinate descent. The number of HALS iterations was set to $10^{3}$. For the RNMF, the oversampling factor was 20, and 2 subspace iterations were used for the QB decomposition.


The same oversampling parameter and subspace iterations were used in both experimental  examples (\ce{TiO2} on \ce{SiO2} and LGPS-NMC). 

An internally developed software package was used to connect existing packages: PDFGetX3 \citep{Juhas2013PDFgetX3}, PyFAI \citep{Ashiotis2015fast}, Hyperspy \citep{Francisco2024hyperspy}, ristretto \citep{Erichson2018Randomized}, and NCEMpy \citep{openNCEM}. This package is available on Github at \url{https://github.com/awwerbro/ePDF}. The notebooks to create the NMF and PCA figures in this paper can be found in a separate repository on Github at \url{https://github.com/awwerbro/fast_mapping_4DSTEM}. Included in this folder is an extra notebook on a purely crystalline, synthetic \ce{WS2} dataset. The \ce{TiO2} on \ce{SiO2} and NMC-LGPS datasets can be found at \url{https://doi.org/10.5281/zenodo.20366781}.

\begin{figure}[!t]
\centering
\includegraphics[width=\columnwidth]{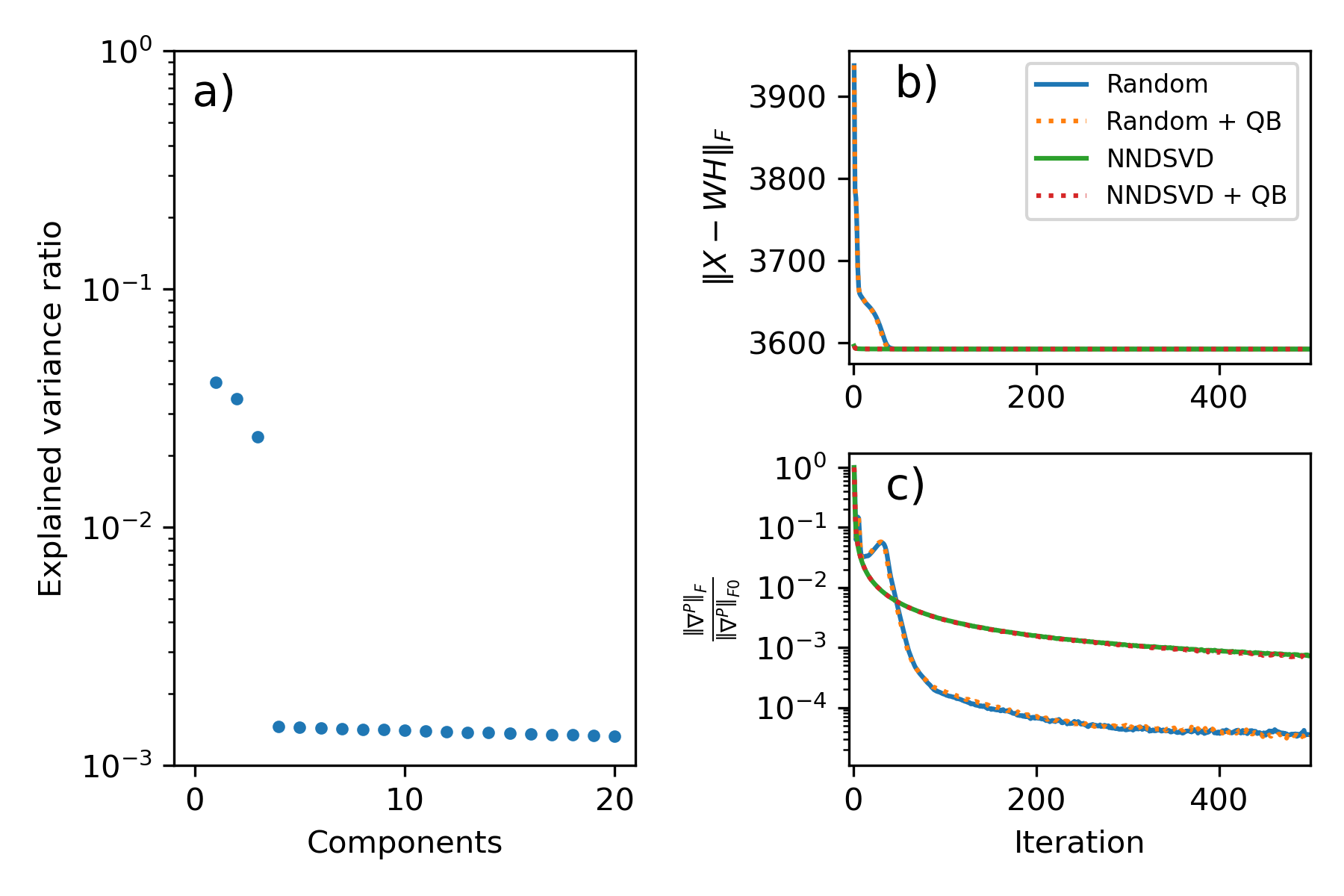}%
\caption{a) Explained variance ratio from a 20-component PCA, showing that, besides the mean component, 3 more components are necessary. b) Loss between factorization and data. The random initialization takes more iterations to converge, but eventually reaches a similar value as the NNDSVDar. The impact of QB decomposition on the convergence behavior is negligible. c) Normalized projected gradient. }
\label{fig_Zr_1}
\end{figure}

\begin{figure*}[!t]
\centering
\includegraphics[width=\textwidth]{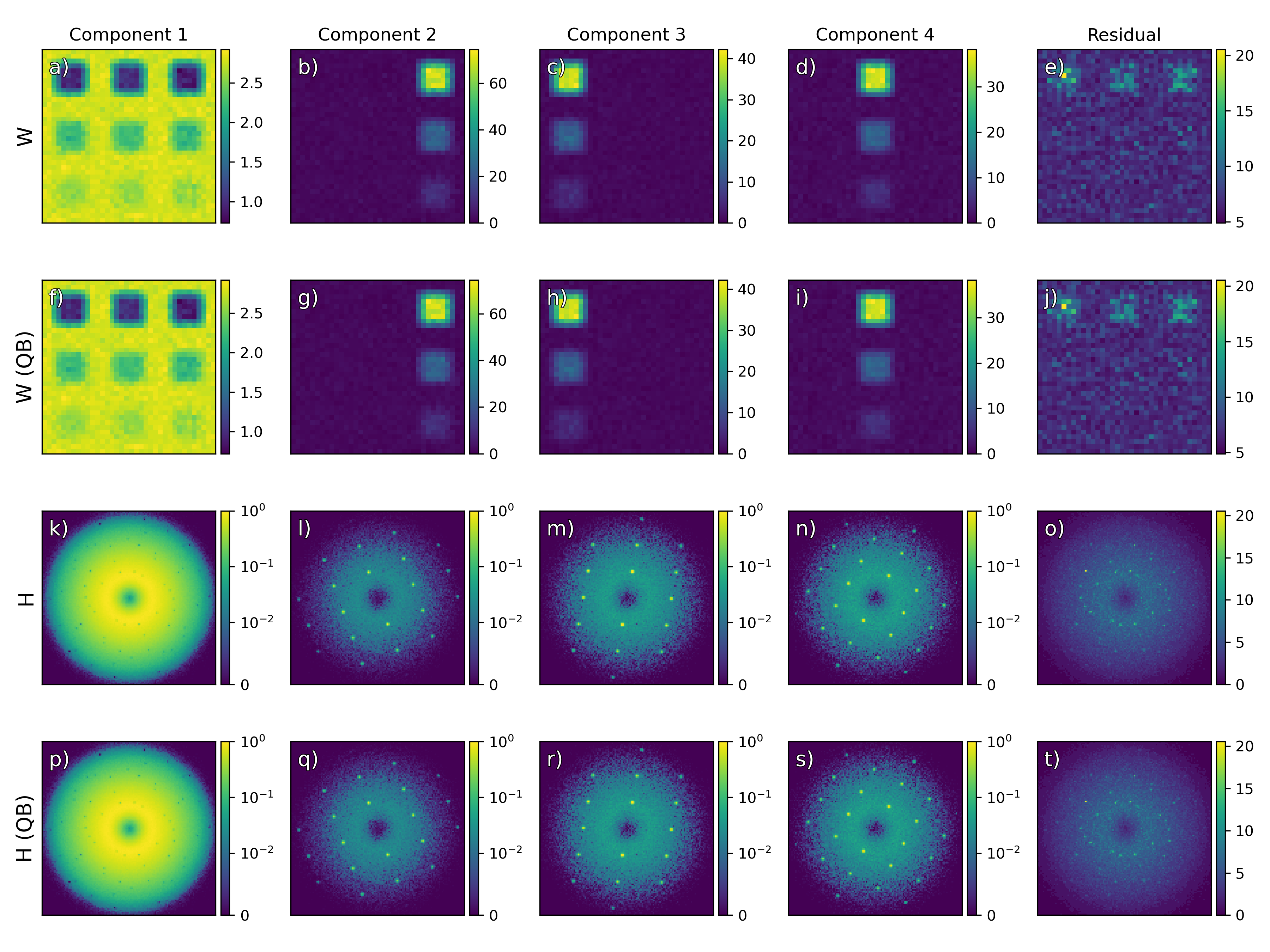}
\caption{RNMF decomposition of synthetic, previously published ZrCuAl dataset with four components, compared to NMF decomposition of the same data. Residuals are compared with the raw data. Components W are normalized, maps H are scaled accordingly. a-d) RNMF maps and e) residual with respect to raw data f-i) NMF maps and j) residual showing the noise level in the data. The difference between these two is shown in Figure S3}
\label{fig_Zr_2}
\end{figure*}

\section{Results}
\subsection{Validation on literature dataset}
As a first mixed phase example, we consider the ZrCuAl dataset constructed by \citep{kimoto2025nonnegative}. The challenge with this dataset is the overlap in reciprocal space between the peaks of the 3 crystalline phases with varying intensity in an amorphous, broad background of the metallic glass. 

We used an exploratory PCA to validate the number of components, although we know there are four components. Figure \ref{fig_Zr_1}a shows how much of the variance is explained by each principal component. It seems that only three components, instead of four, are necessary to describe the complete dataset. We will discuss the reason for this below and we will continue the NMF with four components. The PCA components are reproduced in the SI (Figure S1)

Panel \ref{fig_Zr_1}b shows the convergence behavior of the coordinate descent NMF as a function of iteration. It is clear that random initialization increases the initial error, but also converges to a very similar loss as the NNDSVDar. The convergence behavior between full dataset and QB factorized dataset is extremely similar over these 1000 iterations (of which only 500 are shown for clarity), both for random and NNDSVD initialization. 

Panel \ref{fig_Zr_1}c displays the projected gradient, scaled by the initial projected gradient. Since random initialization started from a higher loss, the absolute value of the initial projected gradient is much higher. At high iterations, close to the minimum, this leads to confusion as to whether the factorization has converged. Again, the full dataset and the QB factorized dataset show extremely similar behavior. 

To show how close the NNDSVD(ar) initialization is to the final factorization, we have included the initial components in the SI (Figure S2). The broad background is almost entirely resolved in the first component although some crystallinity is left, and the other components show a clear preference for the crystalline regions at initialization, despite some overlap. This is a massive improvement from random initialization in which initial components consist of random numbers. Considering the dramatic difference in the initial state between these two initialization methods, it is not appropriate to compare the convergence between different initialization schemes when scaling the projected gradient by the  initial projected gradients.

Figures \ref{fig_Zr_1}b and c also highlight the similarity between analyzing the entire dataset and the QB factorized dataset over 1000 iterations. However, these panels do not indicate the time required to execute these factorizations. To construct Figure \ref{fig_Zr_1}b, we calculate the loss $||X - WH||_F$ at every iteration. Performing this loss calculation at each iteration would be a waste of resources for code running in production. When we do not calculate this  loss at each step, running 1000 iterations of the NMF algorithm on the full dataset takes 600 s on the  (a consumer laptop with AMD Ryzen 9 8945 HS and 32 GB ram). The QB decomposition and 1000 iterations on the reduced matrix take in total 30 s on the same hardware with identical settings. This is a 20-fold acceleration, with (nearly) identical results, as indicated by the losses (Figure \ref{fig_Zr_1}b), projected gradients (Figure \ref{fig_Zr_1}c) and final factors and weights (Figure \ref{fig_Zr_2})

Using NNDSVD initialization and refinement for 1000 iterations, the real space distribution of the four components and residual are plotted in Figure \ref{fig_Zr_2}a-e (NMF) and f-j (RNMF). The reciprocal space fingerprints and residual are plotted in Figure \ref{fig_Zr_2}k-o (NMF) and p-t (RNMF). The difference spectra, which show a negligible difference between the NMF and RNMF results, are included in the SI (Figure S3). The first and second rows (a-j) show the spatial distribution of the different components, whereas the third and fourth rows (k-t) show the reciprocal space 2D diffraction pattern for each component. The first component is the amorphous ZrCuAl matrix that embeds the crystalline particles. The second, third and fourth components show the three different crystal orientations. 

The residuals (Figure \ref{fig_Zr_2} e, j, o, t, last column) are not zero, and as plotted appear to show high error in some pixels. This arises because of our choice to plot the absolute error rather than the fractional error, combined with the fact that the noise in this dataset is synthetically constructed to follow an ideal Poisson distribution. As such, regions with higher peak intensities show a higher deviation from the mean and thus a higher residual. We note here that we are employ a Frobenius norm, which has the underlying assumption that the noise is Gaussian. This is not considered an issue in the literature, as in real experimental (non-synthetic) examples, the count rates will be much higher, converging to Gaussian noise. 

From the NMF components, we can now infer where the additional PCA component comes from during our initial exploratory analysis. The ZrCuAl amorphous phase is present in every pixel and will thus mainly be absorbed into the mean value. This is shown in the SI (Figure S1), where we plot the PCA components. From these plots in Figure S1, it is also clear that the PCA components do not offer the same level of physically interpretable insights as the NMF components.  

From this test on synthetic data, we show that convergence in NMF is defined in a somewhat arbitrary way, depending on a user-defined, initialization-dependent cutoff. We also confirm that NNDSVD(ar) initialization gets us closer to the final factorization, with the benefit of requiring fewer iterations in production. Finally, QB decomposition can effectively compress a large dataset, leading to much faster iteration through the same number of cycles. 

\begin{figure}[!t]
\centering
\includegraphics[width=\columnwidth]{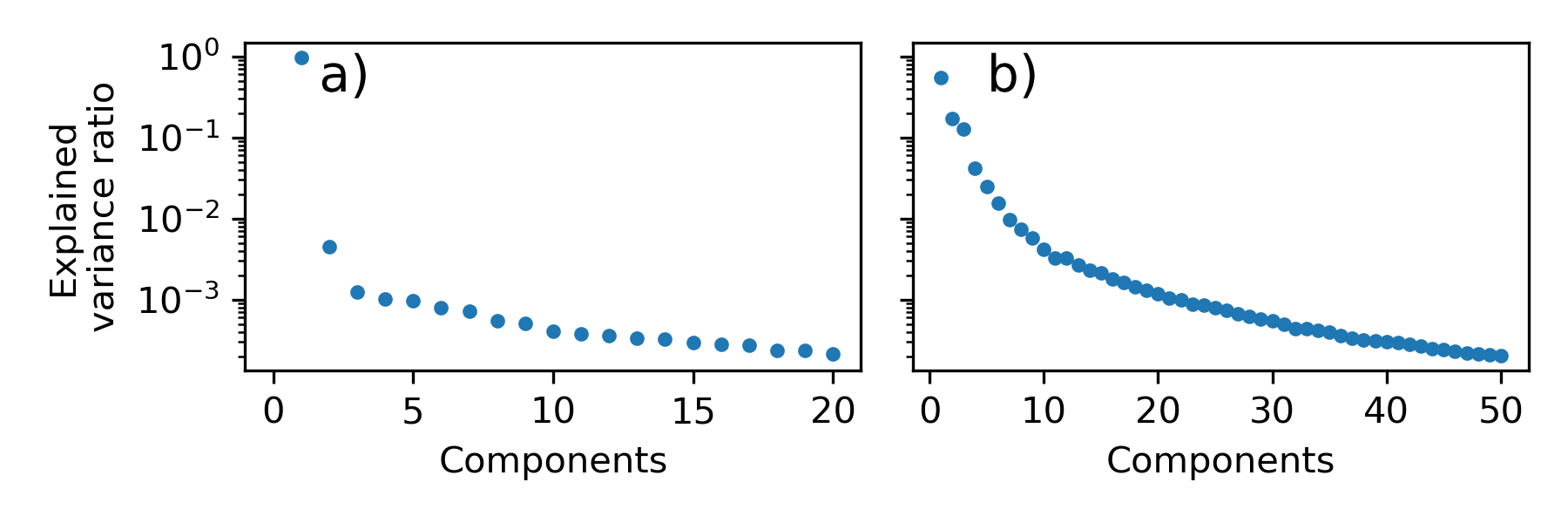} 
\caption{PCA Explained variance ratio for a) the \ce{TiO2} on \ce{SiO2} dataset and b) the NMC-LGPS dataset. }
\label{fig_EVR} 
\end{figure}

\begin{figure*}[!t]
\centering
\includegraphics[width=\textwidth]{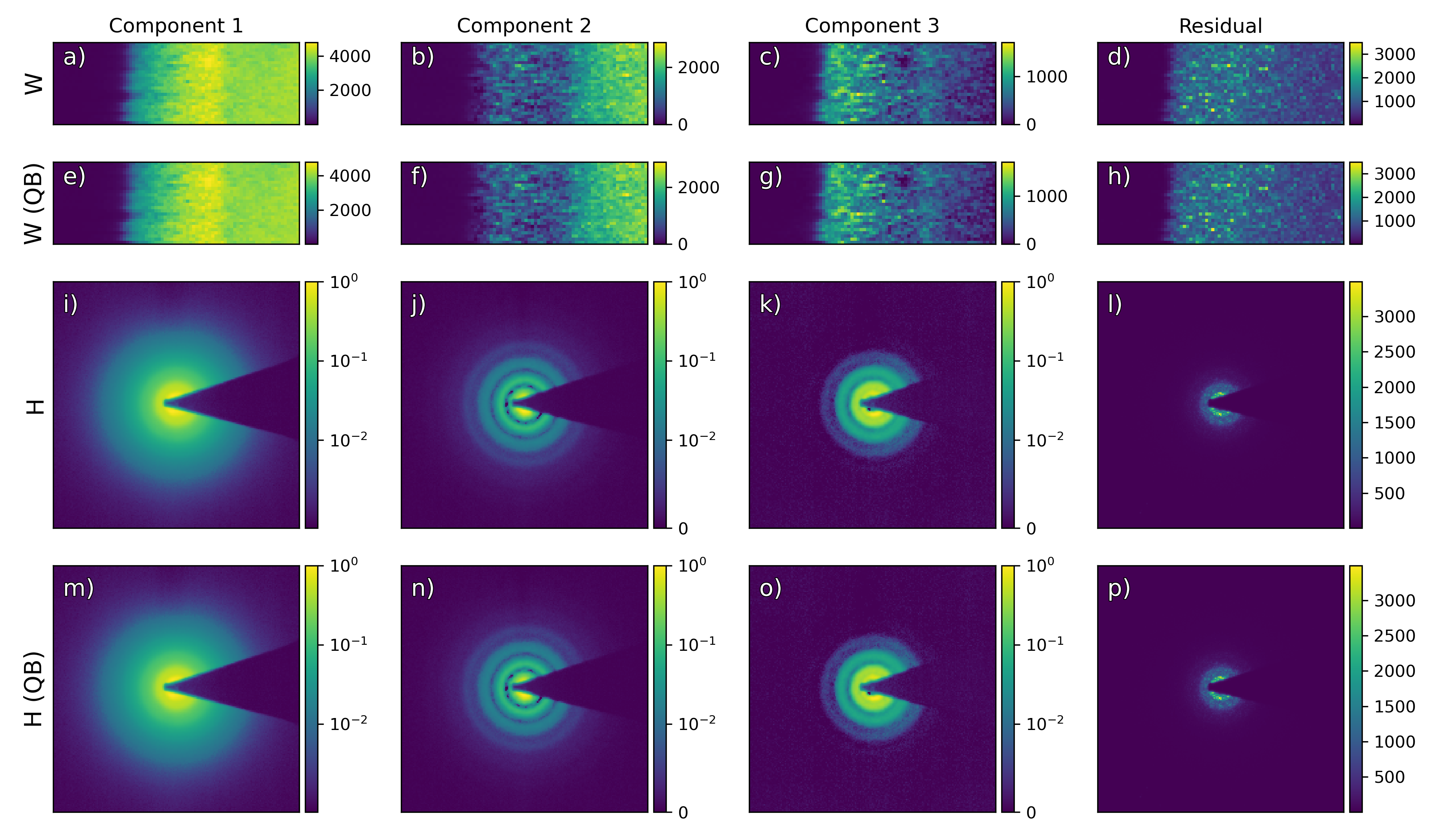}
\caption{RNMF decomposition compared to NMF decomposition of a real world sample: ALD \ce{TiO2} on a \ce{SiO2} sphere. a-c spatial maps of the 3 components as determined with NMF d) residual of the spatial maps with the collected data. e-g) spatial maps of the 3 components as determined with RNMF h) RNMF residual with the collected data. i-k) differences between NMF and RNMf components (see text) l) differences between the residuals. m-o) and q-s) are the reciprocal space fingerprints of the different phases as obtained with NMF and RNMF, respectively, while p) and t) are the residuals with the collected data. panels u-x) show the difference in reciprocal maps. Figure S6 displays the differences between NMF and RNMF components.}
\label{fig_Ti_1} 
\end{figure*}

\subsection{\ce{TiO2} on \ce{SiO2}}
Experimental 4D-STEM data on amorphous materials was also used to test our approach. The first experimental dataset was collected at the edge of a \ce{SiO2} sphere coated with ALD \ce{TiO2} (Figure \ref{fig_1}a). To determine the linearity and number of components, exploratory PCA was performed again (Figure \ref{fig_EVR}a). Besides the mean, two principal components allow to explain most of the variance (Figure S4). Principal components of higher order encompass a very similar amount of the data, only describing noise. 

From Figure \ref{fig_EVR}a, using three components was deemed sufficient to explain our data, and the results of the NMF analysis of the 4D-STEM data for the \ce{TiO2}-on-\ce{SiO2} material system are presented in Figure \ref{fig_Ti_1}. A similar convergence behavior was observed as in the previous example (random versus NNDSVDar, with and without QB). This can be explored through the attached notebooks and seen in Figure S5. 1000 coordinate descent iterations without QB decomposition took about 83s on the computer hardware used in this work, while performing QB decomposition and 1000 coordinate descent iterations took less than 3s. 

The first component from the (R)NMF decomposition is depicted in Figure \ref{fig_Ti_1} (a, e), where Figure \ref{fig_Ti_1} (i, m) shows the corresponding 2D diffraction pattern. Based on the map in Figure \ref{fig_Ti_1} (a), this component spatially aligns with the \ce{TiO2} coating. Because the \ce{SiO2} particle is spherical, as the beam position moves from the surface of the particle toward the bulk, the beam will initially pass through a pure \ce{TiO2} shell. Then, as the beam rasters further into the bulk, the beam will pass through both the \ce{TiO2} shell and the \ce{SiO2} particle; therefore \ce{TiO2} is expected to be present everywhere on the particle. Hence, we identify the 2D diffraction pattern in Figure \ref{fig_Ti_1} (a, e) as the reciprocal signature of \ce{TiO2}. This signature appears relatively featureless with no characteristic diffraction rings. However, this is mainly because of its high intensity. Since this phase makes up the majority of the probed volume, this component seems to have absorbed the background signal (inelastically scattered), too. 

\begin{figure}[]
\centering
\includegraphics[width=\columnwidth]{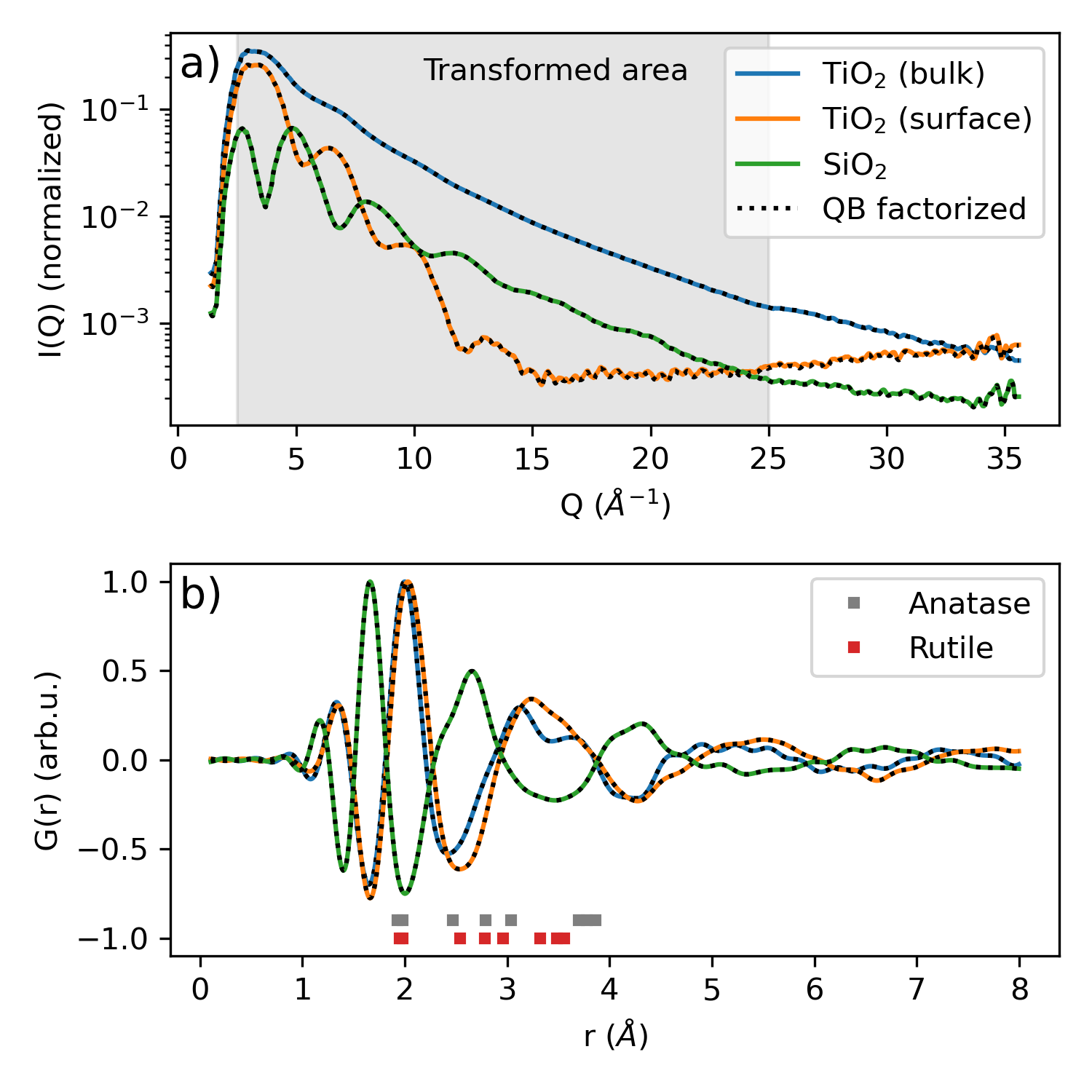}
\caption{PDF analysis of RNMF components including (a) azimuthally integrated diffraction rings (components W) where the grey box indicates the data that was used to construct the PDFs and (b) normalized PDFs (G(r)) calculated for each component from the data in panel (a). There is a marked difference between \ce{SiO2} and the \ce{TiO2} components, but also \ce{TiO2} shows subtle differences between bulk and surface components, indicating a lower density at the interface. Expected peak positions for the crystalline rutile and anatase phases of \ce{TiO2} are indicated.}
\label{fig_TiO2_gr}
\end{figure}

The second component in Figure \ref{fig_Ti_1} (b, f, j, n) corresponds to the \ce{SiO2} core based on the map in Figure \ref{fig_Ti_1} (d). We note that some excess signal can be seen in the spatial map within the outer shell where only \ce{TiO2} is expected to be present. Interestingly, the third component in Figure \ref{fig_Ti_1} (c, g, k, o) seems to be located at the \ce{TiO2} surface. 

The residuals are displayed in Figure \ref{fig_Ti_1} (l, p), and the spatial map of the residual $r_\text{real}$ is displayed in Figure \ref{fig_Ti_1} (d, h). Some streaking artifacts are visible. We have no explanation for these. Difference spectra for the factorization of this dataset to allow comparing the results with and without QB decomposition, are available in the SI.

To highlight the interpretability of NMF components, and the fidelity even when QB decomposition is used to pre-process the data, pair distribution function (PDF) analysis was performed on these components. Each 2D diffraction  component depicted in Figure \ref{fig_Ti_1} was integrated azimuthally as presented in Figure \ref{fig_TiO2_gr} (a), and PDFgetX3 software employed the 1D diffraction data between Q=2.5\AA$^{-1}$ - 25\AA$^{-1}$ to obtain a density-scaled reduced pair distribution function $\frac{G(r)}{4\pi\rho_0}$  for each component, as depicted in Figure \ref{fig_TiO2_gr} (b). The PDFgetX3 software models the scattering background as a polynomial, and values of the `rpoly' parameter were 1.78, 1.84 and 1.74, for \ce{TiO2} bulk, \ce{SiO2}, and \ce{TiO2} interface respectively (See Figure S7). These values are within reasonable bounds.

Looking at the actual peaks in G(r), the obtained G(r) functions for each component match expectations based on literature for our expected assignments for the different components to \ce{TiO2} and \ce{SiO2}. In vitreous \ce{SiO2} (\cite{Mozzi1969structure}) the coordination of Si is four-fold with Si-O bonds at 1.62 \AA, the O$\cdots$O distance is at 2.65 \AA,  and the Si$\cdots$Si distance is at 3.12. The 2nd coordination spheres include Si$\cdots$O at 4.15\AA, and the superposition of  O$\cdots$O  and Si$\cdots$Si at 5.1\AA. Disregarding the first, spurious peak at 1.15 \AA (see below), the maxima of the G(r) obtained in this work are 1.65, (2.35), 2.65, (4.0), 4.31 \AA, with the values within parentheses signifying shoulders. This indicates agreement up to the first/second coordination sphere, where discrepancies may arise because of different angle distributions. 

For \ce{TiO2}, pair coordination distances depend strongly on the crystal/amorphous phase. The first coordination sphere distances range from 1.8 to about 2.3 \AA\, \citep{Mavracic2018Similarity}, depending on the bond angle distribution. Ignoring the first spurious peak at 1.31 \AA, we find bulk peaks at 2.00, (2.8), 3.10, 3.67 \AA \, for bulk \ce{TiO2}, which we can ascribe to Ti-O, (O-O) and Ti-Ti bonds respectively (with the Ti-Ti bonds responsible for the two last peaks). For the surface/interface \ce{TiO2}, we find peaks at 2.02, (2.90), 3.22 and 3.53 \AA, where the longer pair distances for each of these peaks signify that the surface/interface phase is less dense than the bulk \ce{TiO2}. 

While we can explain the ascribe the PDF peaks above 1.5\AA\, to the presence of  Si-O, O$\cdots$O, and Si$\cdots$Si pair distances (and Ti analogs) within these samples, no atomic pair is known for the composition of this sample explains the presence of a peak between 1 and 1.5 \AA\, in each PDF. One potential candidate could be O-H bonds, however the O-H pair distance in Si-O-H is 0.969 \AA\,\citep{McCarthy2008laboratory} with similar values for Ti-O-H. Additionally, the scattering cross-section of H is low.  Another feasible atomic pair could be Ti-Cl bonds from residual Cl originating from the \ce{TiCl4}  precursor, however Ti-Cl bonds have a length of 2.18 \AA \citep{None2020MaterialsTiCl4}. Instead, we interpret these peaks as artifacts of the data processing. To substantiate this interpretation, Figure S7 shows in more detail the effect of  user-selected background modeling paramers on the resulting PDF. It is clear that the location of the first peak is highly dependent on the background model in all cases, while the other peaks remain constant This is an indicator that this first peak arises as a Fourier transform artifact.

The interatomic distances of the anatase and rutile phases are indicated as points within Figure \ref{fig_TiO2_gr}b. The amorphous ALD film exhibits higher pair distances for all the key peak positions relative to the anatase and rutile phases (e.g. consider the second peak). We attribute this to a lower material density, where increases in the first and second coordination sphere pair distances indicate an expansion in the material volume and therefore a lower density. It is known that the density of thin ALD films is generally lower than their PVD/larger scale counterparts and is temperature-dependent \citep{Piercy2017Variation}, \citep{DeCoster2018Density}. The incorporation of H, leading to non-bridging oxygens is one potential explanation for this phenomenon, where the coordination of H to O will lengthen the adjacent Ti-O bonds\citep{Jasim2021Cryo}. 

We note that the presence of a distinct surface \ce{TiO2} phase is not necessarily surprising - surface reconstruction has been known for decades \citep{schlier1959structure}. However, such reconstruction typically involves densification of the upper atomic layers, whereas here it seems the surface phase is of lower density and spans multiple nanometers (tens of atoms). Some earlier work on ALD \ce{Al2O3} and \ce{TiO2} suggests that, because the material is deposited in a layer-by layer fashion, surface structures are retained in the bulk of the material \citep{etinger2017density}. However prior work has found that for \ce{TiO2}, Ti-O bonds are longer near the surface and more \ce{Ti^{3+}} was observed \citep{rich2021retention}. Additionally, our group has previously observed a lower density surface structure of only a few \AA ngstrom thickness at the surface and interface of ALD \ce{Al2O3} on CNTs. \citep{Young2020Probing}, \citep{Paranamana2022Measuring} 

From EDS measurements, we observed the presence of trace Cl in the \ce{TiO2} film, seemingly below the surface, in the phase we described as `bulk' (Figure S8). The absence of Cl near the surface may suggest post-deposition release of (H)Cl from the film, which may contribute to a less dense surface phase. However, the Cl signal is weak and may also be below the detection limit in the outer shell of the sphere, where the sampling volume is lower. It has also been reported that over significant timescales, amorphous \ce{TiO2} films (deposited with a different precursor) can crystallize into the anatase phase \citep{wooding2022transformation, Wooding2024Crystalline}. The early stages of this transformation may be evident here. Another  explanation may originate from the impact of the spherical surface geometry on temperature-induced strain, where post-deposition differences in thermal contraction could lead to different strain levels in the material at different radial positions, producing different pair distances. Although isolating the exact origins of the lower density surface phase for ALD \ce{TiO2} is beyond the scope of the work presented here, the ability of (R)NMF to rapidly isolate this surface component with the highest possible signal to noise ratio represents a promising avenue for future studies.

In addition to physical insights derived from computing the PDF, the calculation of the PDF also allows us to examine how ill-defined \textit{convergence} is during the RNMF decomposition. In Figure S5 the Frobenius norm of the difference between data and decomposition seems negligible already after 500 iterations, yet in Figure S9 the azimuthally integrated intensity of the third component continues to change with further iterations. The diffraction intensity stabilizes after about 100000 iterations (determined visually) in Figure S9, while the pair distribution function seems to stabilize after only 1000 iterations, and peak positions remain generally constant thereafter. Figure S10 demonstrates that with QB factorization, 200000 NMF iterations can be completed in the same amount of time for which conventional NMF achieves 5000 iterations, and the initial time penalty of the QB factorization becomes negligible after about 10 iterations for this data.

In total, we demonstrate with this experimental example that  RNMF decomposition helped us rapidly understand that (1) \ce{TiO2} exists in a bulk and interfacial phase within the ALD layer, (2) the interfacial phase of \ce{TiO2} extends a depth of \textgreater 10 nm into the outer surface of the \ce{TiO2} layer, and (3) the interfacial \ce{TiO2} phase has expanded bond lengths consistent with a lower material density relative to bulk \ce{TiO2} structure, in accordance with earlier reports.

\begin{figure*}[!t]
\centering
\includegraphics[width=\textwidth]{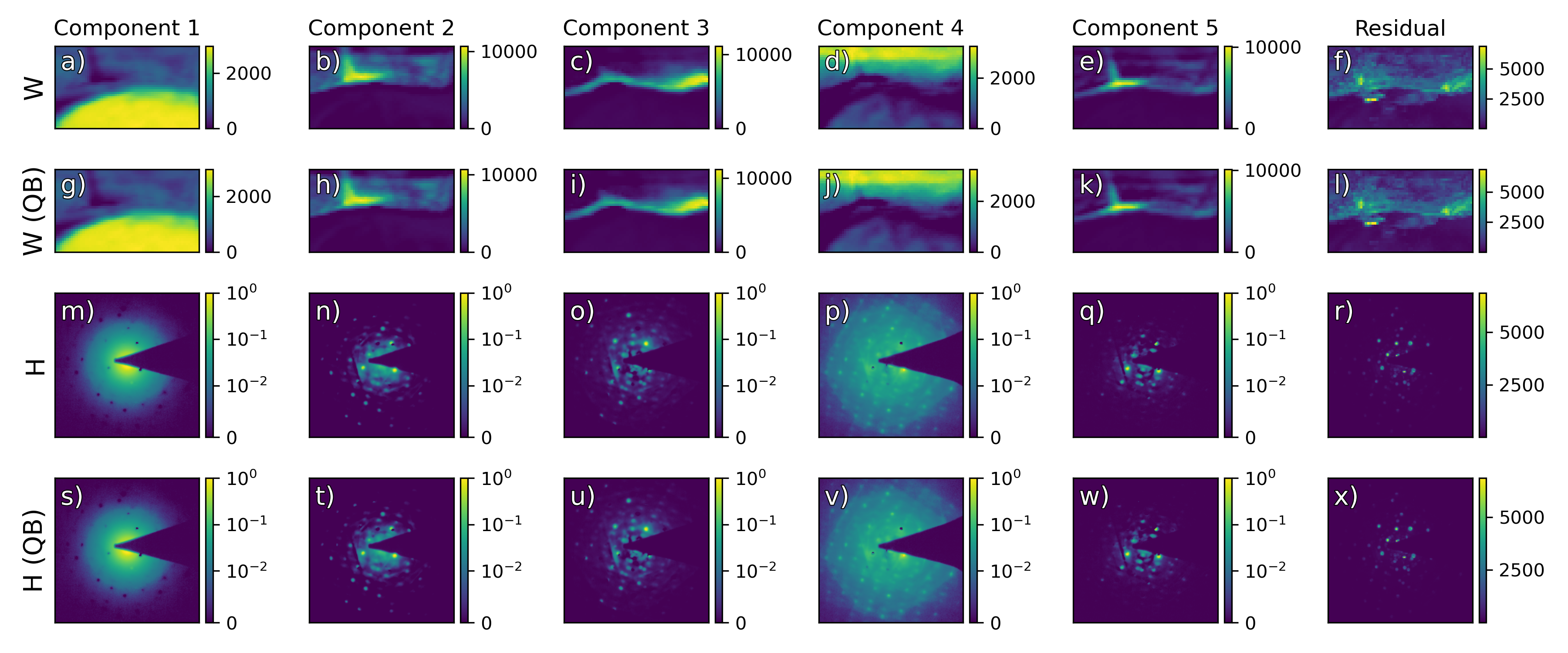}
\caption{RNMF decomposition compared to NMF decomposition of an LGPS-NMC interface. a-e) spatial maps of the 5 components as determined with NMF f) residual of the spatial maps with the collected data. g-k) spatial maps of the 3 components as determined with RNMF l) RNMF residual with the collected data. m-q) and s-w) are the reciprocal space fingerprints of the different phases as obtained with NMF and RNMF, respectively, while r) and x) are the residuals with respect to the collected data. }
\label{fig_BATT_1} 
\end{figure*}

\subsection{NMC-LGPS interface}
For a second  experimental demonstration of the use of RNMF on a more complex, multicomponent system, we revisit a dataset reported in earlier work. In this previous work, we studied interface formation between crystalline NMC and amorphous LGPS. We observed electrochemical reduction of the transition metals in the NMC and migration of oxygen toward LGPS. This effect was stronger when the NMC was chemically delithiated before the two materials were brought into contact. Experimental details can be found in \citep{Paranamana2025Understanding}. In our earlier work, we employed a k-nearest neighbors non-local averaging method to increase the signal-to-noise ratio of diffraction data. However, as discussed in this previous work, challenges arise in separating amorphous and crystalline components, especially in identifying and isolating artifacts from textured domains.

Exploratory PCA revealed an interesting structure of the data (Figure \ref{fig_EVR}b). In contrast to the ZrCuAl and \ce{TiO2} on \ce{SiO2} datasets, there is no clear bend or elbow in the explained variance ratio with an increasing number of components. A reasonable interpretation of such an elbow effect is that the main variations in the dataset are captured the the components prior to the elbo, whereas additional components after the elbow are merely capturing measurement noise. However, such an effect is not observed if nonlinear processes underpin the data: indeed, nonlinear subspaces are not adequately described by either PCA or NMF. We pose that such an ever-decreasing explained variance without clear cutoff signifies that the data lies in a \textit{non}-linear subspace in $n$-dimensional feature space. If this is true, PCA or NMF may not be optimal tools to separate out components and maps. Yet, to the best of our knowledge, no tools exist to deconvolve non-linear subspaces. 

Since there is no clear elbow, we proceeded with NMF analysis using five components: we expect LGPS and NMC to be present, and included three additional components to capture potential interphase material structures. 1000 NMF iterations without QB decomposition took 118 s on the computer hardware used for this work, whereas with QB decomposition the full data processing and NMF analysis took less than 5 s from start to end. Again, very similar components are identified (Figure \ref{fig_BATT_1}), highlighting the benefit of employing QB decomposition to accelerate NMF analysis.

The first component (Figure \ref{fig_BATT_1}(a, g, m, s)) shows the LGPS region. The LGPS seems to be amorphous, and homogenous, as indicated by the lack of second component necessary to reconstruct this region. Strikingly, the diffraction pattern becomes very low/zero where the peaks for lower-order diffraction of other components are expected. This is visible as `holes' in the spectrum despite our lack of sparsity requirements. We attribute these holes to the difference between the scattering mechanisms and due to the lack of pure pixels.

The second component in Figure \ref{fig_BATT_1}(b, h, n, t), third component in Figure \ref{fig_BATT_1}(c, i, o, u) and fifth component in Figure \ref{fig_BATT_1} (e, k, q, w) are located at the NMC surface region. Bragg peaks are visible in each of these diffraction spectra, at similar positions, yet with different intensities, explaining why they are distributed into different components by the NMF algorithm. The origin of these shifts can be explained by variable material orientation relative to the electron beam: as the beam moves across the particle, the orientation of the ordered domains in the material will change with respect to the beam, leading to off-zone axis conditions and amplification of different diffraction spots dependent on the locations. These relative peak intensity differences are interpreted as different components by the (R)NMF algorithm.


The fourth component in Figure \ref{fig_BATT_1} (d, j, p, v) is mapped where the NMC is thicker and contains peaks and Kikuchi lines originating from multiple scattering. 

The residuals in  Figure \ref{fig_BATT_1} (f, l, r, x) are far from perfect. We attribute this to combination of Poisson noise in the peaks, the use of a low number of components, and lack of clear elbow in Figure \ref{fig_EVR}b. We note that the residuals are equivalent when QB decomposition is performed vs. is not performed, as depicted in Figure S11.

This highlights one major limitation of NMF and PCA. When the linearity requirement is violated an 'infinite' number of components can be used to explain the differences in spectra. Examples of such situations include strain or stress gradients present in the material, off-zone axis conditions, thickness effects such as multiple scattering, or a deviation from the kinematic approximation. A remark here is that strain in and of itself is not a problem for NMF, yet strain could lead to nonlinearity. Nevertheless, interesting maps can be obtained. In the github repository with all the analyses for this work (accessible at \url{https://github.com/awwerbro/fast_mapping_4DSTEM}), we have included an analysis of an existing, strained \ce{WS2} dataset, allowing for linear decomposition. This dataset was not included in the main text for reasons of scope and brevity.

\section{Discussion}
Allen \textit{et al.} demonstrated the use of PCA and NMF for grain mapping on a very large dataset of gold-palladium nanocatalysts \citep{Allen2021Fast}. In their work, PCA analysis on a full dataset took 55 s, while the NMF took 44 h (MATLAB implementation). On a smaller dataset (1 GB), 4 minutes of PCA time and 40 minutes of NMF were reported. Due to the faster processing PCA was considered the best candidate for a first-pass analysis, despite NMF offering direct interpretation and better unmixing. 

\subsection{Advantages of NMF}
In this work, we propose that it \textit{is} feasible to employ NMF directly on raw 4D-STEM data. The justification is that NMF analysis is akin to constructing an idea, post-measurement, `virtual detector' which works equally well for crystalline, amorphous, and mixed datasets. We demonstrate this through three examples: a ZrCuAl glass with 3 different crystallite orientations, an amorphous \ce{TiO2} layer on a \ce{SiO2} particle, and an LGPS-NMC battery interface. We note also that virtual detectors assume linearity and pure pixels to navigate 4D measurement space. 

NMF offers two distinct advantages over PCA: (1) In real space, the component intensities can be interpreted directly as amorphous and/or crystalline phases, mapping them to positions in real space. One pixel in real space can contain an arbitrary volume of each identified phase. (2) Meanwhile, in reciprocal space, the NMF algorithm distills the available data down to its purest component forms, minimizing sampling noise in the process. In this way, high quality diffraction data are obtained for each identified phase. Furthermore, NMF employs as much collected data as possible, effectively allowing us to present a complete 4D dataset in a meaningful and interpretable way. 

Computational demands for standard NMF exceed (some) practical limits on 4D-STEM datasets, in particular the amount of iterations necessary to reach convergence. We demonstrate that the convergence time can be greatly decreased by (1) using QB decomposition as a preprocessing step to reduce the sample space, and (2) using NNDSVD as an effective initialization. 

In this discussion, we question the de facto definition of convergence in previous work based on  normalized projected gradients and an arbitrary cutoff. While for smaller datasets, it is of low impact to run (for example) 10000 more iterations to reach a desired cutoff, in the context of large datasets , increasing the number of iterations at that scale quickly becomes unfeasible. Reducing the number of iterations for such larger datasets appears customary in the literature. Kang \textit{et al.} mapped surfaces of metallic glass and used a MATLAB implementation of NMF to discern two intermixed sub-phases \citep{Kang2024Mapping}. The authors did not mention how long their factorization took, but 1000 iterations were necessary before convergence. Uesugi \textit{et al.} used 4D-STEM to probe 100×100 spatial positions on \ce{TiO2} nanosheets with 128×128 pixels in diffraction space, creating a data matrix of 650 mb \citep{Uesugi2021Non}. Using Digitalmicrograph scripts, they mention 20 iterations to be sufficient to reach separation into interpretable components. Kimoto \textit{et al.} \citep{Kimoto2024Unsupervised} used a similar approach on 3364 spatial positions, 128×128 pixels in diffraction space. Here, convergence took only 100 steps after 20 minutes of runtime. 

The number of iterations selected by the user seems to be tacitly dictated by computational needs. Here, we show that for a low number of iterations, the extent of convergence will depend on the initialization. For quick mapping and visual inspection, this is likely sufficient. However, for more detailed analysis, convergence should be explicitly examined. 

In Figure S9, we show this explicitly for the case of \ce{TiO2} on \ce{SiO2}. The first two components converge relatively quickly, but the third component needs up to 10000 iterations. On the other hand, we verify that the PDF converges around 1000 iterations. Here, QB decomposition allows us to run a high number of iterations, ensuring convergence. Running 200000 iterations took only 215 s (Figure S10). 

\subsection{Other preprocessing approaches}
A simple preprocessing step to reduce the data size is azimuthal integration. However, this becomes a problem when mixed amorphous/crystalline materials are studied, since the presence of Bragg lattice peaks can convolute the reduced pair distribution function G(r) and orientation/strain/stress information is lost upon integration. Other work uses NMF on the reduced pair distribution function G(r) itself. A cloud-based web application nmfmapping is available \citep{Thatcher2022nmfMapping}, and a two-phase algorithm for NMF on streaming PDF data was developed by Gu \textit{et al.} \citep{Gu2023Fast}. Liu \textit{et al.} reported on NMF mapping on time data \citep{Liu2021Validation} and also 4D-STEM data was analyzed this way \citep{Rakita2023Mapping}. In each of these cases, azimuthal integration is performed before any other analysis is done, erasing Bragg-based lattice information in the process. Another potential issue the authors of these studies bring forward is that G(r) is not positive everywhere, hence the curve is shifted upwards as a part of the preprocessing routine. Linearity of the signal is necessary for NMF to work well, while shifting could move the data in the positive orthant, but not necessarily within a convex cone necessary for NMF. Furthermore, Mu \textit{et al.} have argued against the use of NMF on pair correlation functions in favor of independent component analysis (ICA) \citep{Mu2021Unveiling}. Here, we are able to circumvent this discussion by performing NMF on raw, 2D diffraction data, which is count-based, linear, inherently non-negative and preserves Bragg peak positions. 

\subsection{Pitfalls and caveats}
Although QB decomposition offers a dramatic drop-in acceleration with minimal tradeoff in performance, the use of NMF on 4D-STEM datasets can be tricky and interpretation of the results can be subtle. In the following, we summarize some key insights from our work. 

\textbf{Input data}
Blind hyperspectral unmixing (or NMF) becomes much easier when \textit{pure pixels} are present \citep{gillis2013fast}, as they form natural boundaries to the convex cone in the positive orthant that encompasses the entire dataset. This is not necessarily the case in 4D-STEM. For example, in the ZrCuAl example, the amorphous background is present everywhere. As such, the crystalline regions can consist of $H_\text{crystal} = H_\text{bragg} + x \cdot H_\text{background}$ when the weighting of the background is adjusted. The lack of pure pixels is one reason for the dimples in the diffractograms of amorphous components, as shown in ZrCuAl (component 1) and NMC-LGPS (component 1) examples. The approach developed by \citep{kimoto2025nonnegative} is specifically aimed at alleviating this issue by adding additional constraints to the components.

Another issue is that different beam-material interaction mechanisms will exhibit different linear responses, especially when moving away from the kinematic approximation. For example, in other examples (not included in this work) we have noticed that the central beam can become its own component, arising from spatially dependent thickness changes in the material. 

\textbf{Algorithm}
As mentioned above, determining the \textit{correct} number of components in a given dataset is an open problem. Here we used the assumption of linearity to justify an exploratory PCA analysis in order to estimate the number of components. This relies on PCA yielding a fast and global (yet uninterpretable) optimum. The fact that PCA centers its component around the mean then offers justification to add one more component.

\textbf{Interpretation}
Two different grains with the same crystal structure will produce different signatures and should be treated as such: it should be straightforward to classify these patterns as different components. However, as a given crystal grain experiences stress/strain and/or orientation dependent scattering at its edges, the diffraction pattern will go through a continuum of deformations, allowing one to add ever more components depending on the data available. Clustering approaches (as in \citep{Kimoto2024Unsupervised}) could provide relief here, although they should also be used with care: machine-learned, distance-based clustering will not be able to distinguish different projections of the same unit cell. This means that our interpretation of the amorphous \ce{TiO2} phase into two components could be viewed as a continuous shift between the two phases as a function of depth rather than two distinct phases. For amorphous materials, this separation goes smoother than for crystalline materials, as the features (rings) move more gradually than the much smaller Bragg diffraction spots characteristic for crystalline materials. For crystalline material, Vegard's law describes the same effect. A similar argument can be made for the NMC-LGPS interface, where the effect is significantly stronger because of the discrete nature of the Bragg peaks. 

One could reason that (R)NMF hence is not useful for the analysis of 4D-STEM data and should be reserved for measurements where clear and discrete components are present. However, we argue that visualizing and understanding the minute changes in the structure of the material using (R)NMF is useful: after all, the surface \ce{TiO2} is structurally different from the bulk, as are the NMC-LGPS interface components. (R)NMF could be used to discretize continuous variables such as stress, strain and sample-beam interactions, effectively providing interpretable touchstones for these effects. Another approach could be to use NMF to purify interface signatures: a large number of components can be used to identify the unmixed phases and approximate interfacial phases. The unmixed phases could then be subtracted from the dataset to obtain continuously changing difference spectra for the interface.

Just like tolerances (projected gradients), residuals (Frobenius norm of the difference between data and model) are not very reliable to indicate convergence. Because the counts are Poisson-distributed, higher-count regions will have a higher absolute noise. Furthermore, we point to the case of \ce{TiO2} on \ce{SiO2}, where we explicitly show that for the change in the components as a function of iteration, the residual stabilizes after only a few iterations, while the third component continued to subtly evolve for hundreds of iterations.

\textbf{Use on larger datasets}
Aligning with the Jevons paradox, using our approach for 4D-STEM data analysis will make surveying large areas with more pixels and higher resolution in reciprocal space tractable. However, we find that NMF is most useful to compress a moderately-sized dataset into a couple (less than 10) components and maps. Since components will often be spatially correlated, we propose to \textit{divide and conquer} larger datasets, treating them as several spatially separated datasets. We foresee that hierarchical clustering approaches will prove to be valuable in this respect.

\section{Conclusions}
We introduce the use of randomized non-negative matrix factorization to separate components in raw 4D-STEM data. QB decomposition leads to an acceleration of up to two orders of magnitude compared to classical non-negative matrix factorization with minimal induced error. This allows one to choose the number of iterations rationally rather than selecting a minimal number based on computational tractability, and enables phase mapping of amorphous and mixed data within reasonable timescales. We demonstrate this on 3 distinct systems: (1) on synthetic ZrCuAl data, (2) by identifying and mapping two phases of \ce{TiO2} in a 20 nm thick ALD coating, and (3) studying a mixed interface between crystalline NMC and amorphous LGPS. 

Further analysis reveals the presence of a low-density \ce{TiO2} surface phase of ~10 nm thickness on ALD grown \ce{TiO2} in the second experimental example system. We are also able to precisely map the modified surface phases of delithiated NMC in contact with LGPS despite the nonlinearity of the dataset. Using (R)NMF to rapidly distill high volumes of raw 4D-STEM data will enable more rapid discovery and innovation in materials science at a multitude of length scales.


\section{Competing interests}
No competing interest is declared.

\section{Author contributions statement}
A.W. and M.J.Y. conceived the project. A.W. developed the code and conducted the analysis. N.C.P. and A.W. prepared the samples. X.H. collected the microscopy data. A.W. and M.J.Y. wrote and reviewed the manuscript. All authors contributed to the interpretation of results.

\section{Acknowledgments}
A preprint version of this manuscript was previously posted at \url{https://arxiv.org/abs/2507.17068} \citep{werbrouck2025fast}.
The authors thank the anonymous reviewers for their valuable suggestions and critical questions.  A.W. acknowledges support from the University of Missouri Materials Science and Engineering Postdoctoral Fellowship. N.C.P. and M.J.Y. acknowledges partial support from the National Science Foundation Electrochemical Systems Program through Award Number 2219060. M.J.Y. acknowledges partial support from the National Science Foundation Division of Graduate Education NRT Program through Award Number 2243526. 

\bibliographystyle{plainnat}
\bibliography{lib}
\end{document}